\documentclass[preprint, 12pt, a4paper]{elsarticle}
\usepackage[utf8]{inputenc}

\usepackage{todonotes}

\usepackage{amssymb}

\usepackage[hyphens]{url}
\usepackage[hidelinks]{hyperref}
\usepackage{float}
\restylefloat{table}

\journal{}

\begin{document}



\begin{frontmatter}



\title{Měj Přehled / Be informed}


\author[gym]{Michal Bravanský}
\author[vsb]{Jan Platoš}

\address[gym]{Secondary Grammar School of Nicolaus Copernicus, 17. listopadu 526, 743 11 Bílovec, Czech Republic}
\address[vsb]{Department of Computer Science, FEECS, VSB - Technical University of Ostrava, 17.~listopadu~2172/15, 708 00 Ostrava, Czech Republic}

\begin{abstract}
The amount of information available to the general public is enormous, and it is challenging to extract meaningful and reliable content. The availability of news sources and their trustability are the biggest problems for selecting the proper authority. Moreover, news providers' inability to publish good news on social media creates a gap for young people to access trustworthy information. This article describes an automated system that can extract the essential news and publish them in a different form on social media networks.


\end{abstract}


\end{frontmatter}


\section*{Current code version}
\label{}

\begin{table}[H]
\begin{tabular}{|l|p{6.5cm}|p{6.5cm}|}
\hline
\textbf{Nr.} & \textbf{Code metadata description} & \textbf{Please fill in this column} \\
\hline
C1 & Current code version & v1.0.0 \\
\hline
C2 & Permanent link to code/repository used for this code version & \url{https://github.com/MichalBravansky/MejPrehled} \\
\hline
C3 & Code Ocean compute capsule & \\
\hline
C4 & Legal Code License   & MIT \\
\hline
C5 & Code versioning system used & GIT \\
\hline
C6 & Software code languages, tools, and services used & Python \\
\hline
C7 & Compilation requirements, operating environments \& dependencies & \\
\hline
C8 & If available Link to developer documentation/manual &  \\
\hline
C9 & Support email for questions & michal.bravansky1@gmail.com \\
\hline
\end{tabular}
\caption{Code metadata (mandatory)}
\label{} 
\end{table}








\section{Motivation and significance}
\label{sec:motivation}

As the years go by, the general public is more and more overwhelmed by information overload. This might be the most apparent in news coverage, where each person has access to dozens of news sources. Combining this almost infinite feed of information with the increase in fake news, people might have problems finding a reliable and efficient news source. Numerous studies also back up these claims. From a survey published by Gallup/Knight Foundation in the middle of 2020~\cite{first_survey}, it is apparent that over 90\% of people think that there exists at least a small bias in their favorite news sources. Furthermore, children and teenagers are even more vulnerable. From a survey by Robb~\cite{second_survey} in a group of young people from 10 to 18 years, only 44\% of respondents said they could tell fake news stories from real ones, and 30\% even admitted to sharing a fake news story in the past six months on social media.

This article focuses on a single type of news coverage, the one on social media. In the Czech Republic, most news sites still have not found a way to distribute news information on Facebook, Instagram, or another social network platform. Furthermore, their goal is not to publish news stories but to generate clicks on their sites. Therefore in Czechia, many student initiatives trying to change this, mostly by developing non-commercial news accounts, exist. However, these accounts are based on volunteering and can be abandoned at any time. 

This article proposes a new automated system that publishes news on social network accounts and creates a reliable news feed for its followers without any human help. We have developed a complex system that can extract the most important and relevant news from the online news-feeds from most news providers. The system uses various techniques for the proper news selection that utilizes keyword extraction, document clustering, and artificial intelligence. The system is designed to select the most important events for any time frame from history.  

The system architecture is powerfully described in Section \ref{sec:swdesc}. The section also contains a description of the algorithms and techniques used in the system. The Illustrative example is then depicted in Section \ref{sec:illustrative} and the overall impact of the system is summarized in section \ref{sec:impact}. The final Section \ref{sec:conclusion} summarizes the whole article and highlights the most exciting parts.

\section{Software description}
\label{sec:swdesc}

\begin{figure}[htp]
    \centering
    \includegraphics[width=12cm]{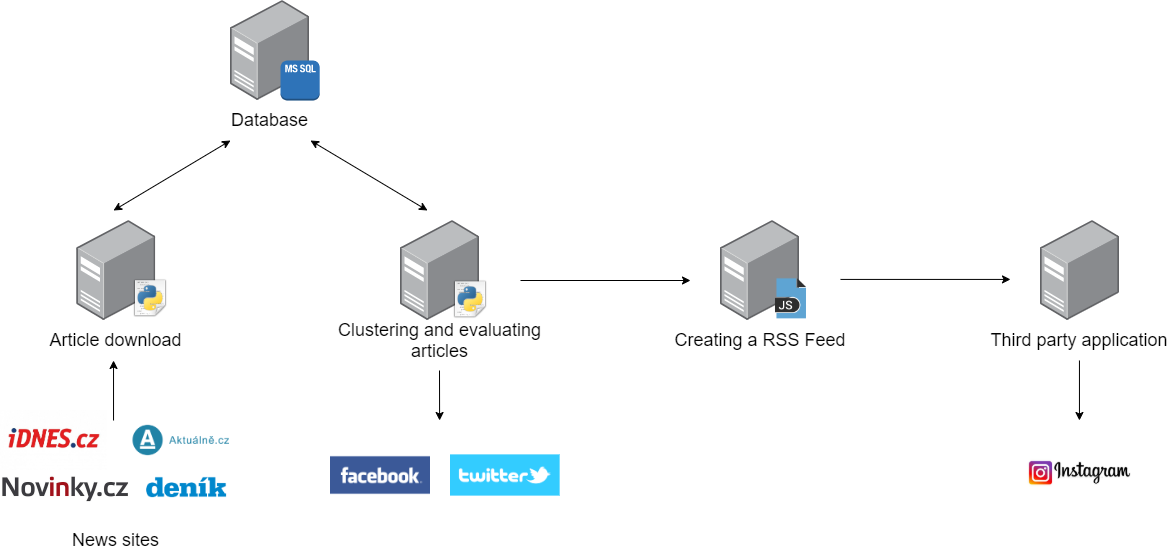}
    \caption{The software architecture of the system}
    \label{fig:diagram}
\end{figure}

The software system is composed of several components. The overall structure is depicted in Figure \ref{fig:diagram}. These components are designed to create a complex system that can be extended using other news sources, different article processing steps, and essential news extraction.

The first part of the software is the news-feed aggregator. The system aggregators are based on the published RSS feed of the news publisher's website. If they come across a new or updated article, they save the database's changes for further processing. Each record from the RSS feed is then accessed on the publisher's website by the article reader system, and all required information is then stored. This information includes the perex of the article, its full text, and the main image. 

The second and even more crucial part of the system is the critical news selection module. This module is responsible for the selection of the most important news from the defined time-interval. The selection process is based on the text vectorization using word-embedding, the clustering algorithm, and the cluster evaluation according to the criteria. The selection process is described in detail in the following section.

\subsection{Dataset preparation}
The excellent text processing system's design requires a good model for the selected language or a large enough dataset. Due to the lack of both mentioned requirements for the Czech language, we decide to create the dataset and the model ourselves. We have created a database of historic Czech news articles from the leading news publishers. The database contains almost a half-million articles dating back to the 2000s. These articles were downloaded using the website scrapping, and all the necessary parts were extracted. 

\subsection{Word Embedding}

Before a FastText Language Model \cite{bojanowski2016enriching} trains on the dataset, the articles need to be tokenized and lemmatized. The article is divided into sentences, and each sentence is tokenized. A Python library Nltk is used for the sentence and word tokenization. Each token is then lemmatized using a library named Majka. If Majka cannot lemmatize a specific word, for example, the last name or a foreign word, it does not alter the word and keeps the original token. FastText is capable of basic lemmatization on its own. Therefore the model can handle these exceptions. Then the stop-words contained in a Czech stop-words dataset are removed. The tokenized sentences are then fed into the model with other additional settings described further.

The recommended number of iteration for the FastText model is between 10 and 30 epochs. Our dataset is composed of articles from different sources with many possibly repeating parts, leading to the model over-fitting. The over-fitting may have an impact on the performance of the model. To avoid over-fitting, we choose just ten epochs.

The recommended vector size for FastText is between 50 and 300. Multiple models were created in this range. Each model then processed the dataset, and a similarity search was performed. To a random article, the most similar articles were found using the cosine similarity. After comparing the results, the model that uses 200-dimensional vectors was the most successful, so we set the dimension to 200 for further processing. The model's learning rate was set to 0.1, and the Window size was set to 5. Other settings were left with default values. These skip-gram properties have little significance in the process of text vectorization. Therefore we did not customize them.

The processing of the real articles that are evaluated is done in the same way as the dataset. The articles are tokenized, lemmatized, stop words are removed, and then vectorized. The first $N$ token is selected from these tokens, described as vectors by the FastText model, and averaged out. After numerous testing with different values of $N$, ranging from 10 tokens to 300, We concluded that the first 50 tokens from the title and the article's content create the most precise vector representations. 

\subsection{Clustering}

The program keeps track of all the articles released in the last 24 hours. This has two effects. First, all the calculations are done on a small corpus of articles, and therefore it is fast. Second, the program should post only about current events. Furthermore, to ensure that the news feed is current and the most informative, the application has to use a time window to select the articles.

After the articles are vectorized, the program creates a similarity matrix. Cosine similarity of the vectors is used to determine the similarity between articles. Each vector is normalized to create the similarity matrix. After that, it is possible to create the product of the matrix with itself transposed. This product is the similarity matrix.

The primary software function is the clustering algorithm. Clustering is done by the modified DBSCAN~\cite{dbscan}. The used algorithm is different from the regular one by not including the neighboring points. By eliminating these points, the cluster is left with fully connected subgraphs for each topic. This means that all the articles inside a cluster are similar to each other. 

Application initializes a  non-directional graph from the Python library NetworkX. Points representing the articles are added to the graph. If the cosine similarity between the two articles is more significant than a specified threshold, the points represented by these articles are connected. The application selects all the cliques from this graph, all the fully connected subgraphs in the graph. This method is faster than using standard DBSCAN because it does not take into account neighboring points. The threshold is set to 0.92. However, it might differ on other datasets.

This method returns a list of all the cliques - fully connected sub-graphs. Therefore, one article can be included in a number of them. To eliminate these duplicity clusters, we first evaluate each cluster and give it it is rating as a number. By this rating, the application sorts all the clusters in descending order. This sorted list is then iterated. If a cluster contains any article, which was already assigned a better score, the cluster gets deleted. This ensures that each article is only associated with a single cluster, and this cluster has the most significant rating available.

The rating is based on multiple properties. The first one is the cluster's size, where we expect a more extensive cluster to be more critical. Another aspect is the diversity of the sources.
The clusters which come from different sources are rated higher than single-source clusters. This ensures that the more critical news is backed up from different outlets and helps to avoid misinformation. Other variables involve the time difference between the releases of the articles and their average length. However, these metrics are secondary, and they are used only if in need of an absolute ranking.

In the end, if an evaluation of the cluster is above the selected threshold, a post about the topic is published on social media.

\subsection{Post generation}
First, it is needed to create an image and a description of a specific topic. For this task, the program selects an article whose vector is the closest to the average vector of all the vector representation in the selected cluster. This means that this article combines the most information from the other articles, and it is not biased. For example, if fake news got inside one of these clusters, the program would not select that specific news because it would not be similar to the other articles on that topic.

This article then represents the whole cluster. The opening image is extracted from the article. This picture is loaded in Python through library PIL and then cropped to achieve a square shape. This cropped image is combined with a frame with multiple colors: blue, orange, or yellow. Each color is then used for a different type of post. The blue frame represents national news, orange international, and yellow essential news. For this purpose, one of these three categories is picked for the article. If the article is national or international is decided by the subcategory to which the text belongs. For the important attribute, the cluster needs to be once again evaluated above a selected threshold.

After a frame is added to the opening image, the selected article's title is written in the bottom left corner. The font used is named Din Condensed. This text can be of two sizes, one bigger for shorter titles and one smaller for longer.

The description is made from the first paragraph of an article. Compared with summaries in RSS Feeds, the first paragraph of an article is usually longer and more informational than the summary. Furthermore, the FastText vector is made from $N$ first tokens of the article. Therefore the first paragraph is partially represented in that vector. This means that this first paragraph is crucial with working with the vector representation, and therefore it is ideal to use it as a post description.

\subsection{News posting on the social platforms}

After creating the image and the description, the program needs to post them. We created three accounts on social media, on Facebook, Twitter, and Instagram for this part of the project. 

\subsection{Facebook}

If an app requires access to Facebook API, it needs to be registered at Facebook for Developers. There the developer signs in with his Facebook account and validates his identity with his ID card. Then the developer can create an application. We created one for Měj Přehled. He selects the required permissions. Měj Přehled only publishes posts. Therefore only publish\_pages and manage\_pages permissions were requested.

After a developer requests permissions, he needs to submit his request, where he describes what he will do with each permission. After the submission, We waited for a week for it to be verified.

The program can then send a POST request to Facebook API with the image, description, and Application Token for verification. This method can publish POST to any Facebook page. However, the program will use it just for one, and that is Měj Přehled.

\subsection{Twitter}
For Twitter, a developer also needs to register his application. However, on this platform, it is much more comfortable. The developer needs to apply for a developer account, where he specifies why he wants to use the Twitter API. After that, we registered the application and again submitted a report about our project. The report was almost instantly approved, and the application could start to use the API.

To make the API more accessible, the program uses a Python library called "twitter." This library allows users to only their credentials and handles all the authorization. The program can also easily publish a new tweet by calling a simple method and passing the text.

Měj Přehled's Twitter posts are a little bit different in comparison with the other platforms. Instead of an image, the tweet contains only the title and the URL of the article. Due to the limited number of characters on Twitter, it is impossible to pass enough information to the post, and the only 
the way how to inform people is by providing an additional source, in this case, the article.

\subsection{Instagram}
In 2018, Instagram shut down its API because of the Cambridge Analytica scandal. As of September of 2020, the Instagram API can still be used only by the selected number of providers. Therefore our application has to output an RSS Feed, which will be passed into a third-party application, which will publish posts to Instagram.

For creating the RSS Feed, we used Google's service Firebase. This service allows us to build apps fast without managing infrastructure. The application will save all posts in a non-SQL database called Firestore and then displays them as RSS Feed. The project will be able to add a new post through a different function.
All backend is made with Node.js. Firestore database contains a collection of all the posts, and each post has a few properties, ImageUrl, PublishDate, and Title. Each of these properties is important. ImageUrl is used to access the image and Title for the description. PublishDate is then used to sort posts.
In the RSS Feed, it is impossible to post a picture's binary. Therefore the program needs to add a public URL to the photo so that the third-party application can access the image. The image will be uploaded to the site ImgBB, which provides free image hosting. In their API, we provide our account token and photo in binary. After the upload, the photo stays public for a month, then gets deleted.

The third-party application we choose is called OneUp. It provides a fast way to publish posts from RSS Feed, and it is the cheapest option on the market. After the Instagram account is connected, and the RSS feed provided,  the app manages everything by itself.

Měj Přehled also adds a set of hashtags to each Instagram post. These hashtags are used to boost the posts' reach and are composed of the hashtags with the largest following in the Czech Republic.


\section{Illustrative Examples}
\label{sec:illustrative}

\subsection{Instagram account}

\begin{figure}[htp]
    \centering
    \includegraphics[width=12cm]{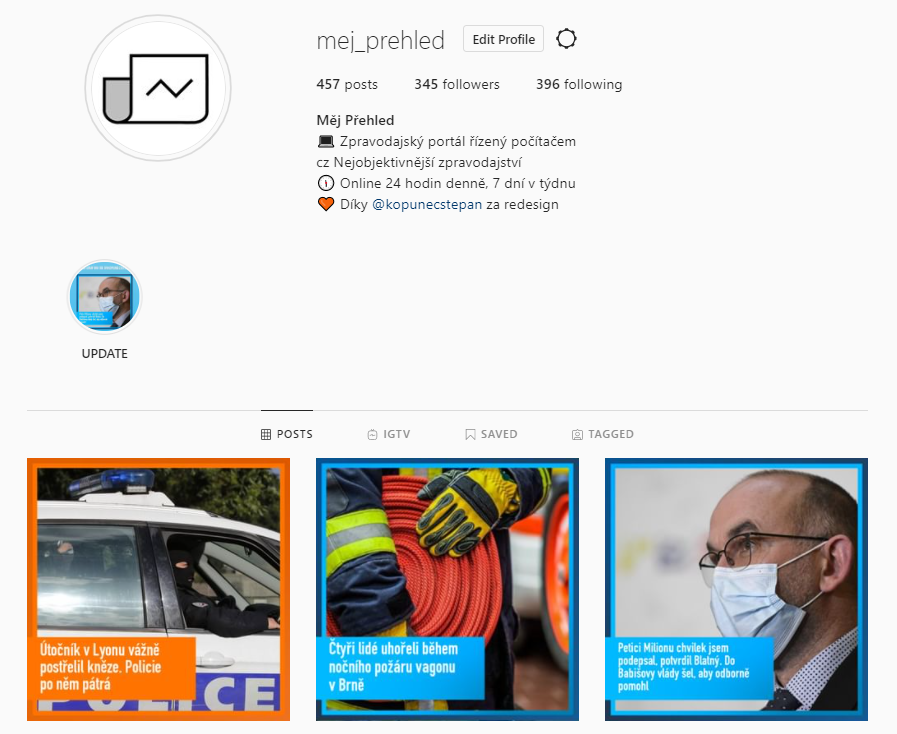}
    \caption{Instagram account}
    \label{fig:example}
\end{figure}


\begin{figure}[htp]
    \centering
    \includegraphics[width=12cm]{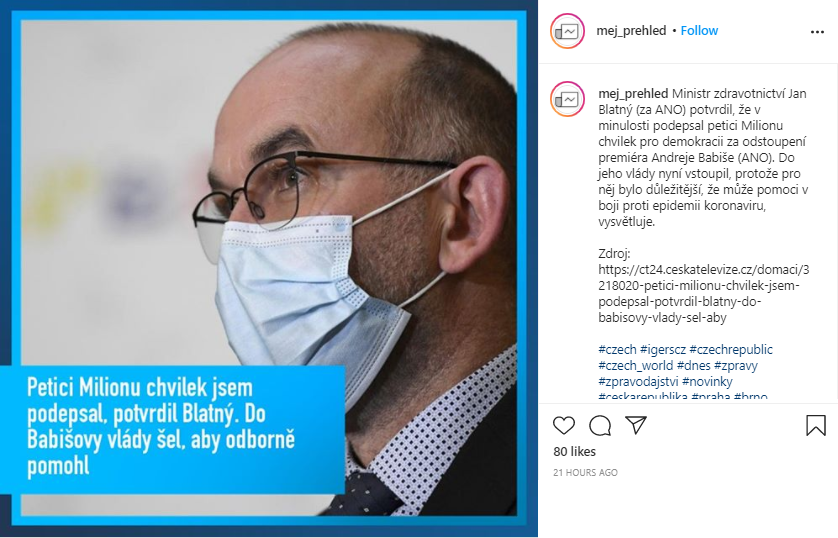}
    \caption{Example of a published post}
    \label{fig:post}
\end{figure}



\section{Impact}
\label{sec:impact}


All accounts have accumulated as of October of 2020 about 1000 followers in total.

This project won Students` Professional Activities in the Czech Republic in the IT category and will participate in the international science fair EUCYS at the end of September 2021.

The basic algorithm was implemented at Seznam a. s., one of the largest Czech IT corporations. There it is used to filter similar articles and help to find the most important ones.

This algorithm also runs inside an app called newskit, a news aggregator with over five thousand downloads. There it selects the most crucial news articles and suggests them to users.

Furthermore, the program was mentioned in the magazine Euro 20 under 20, where reporters select the most interesting projects made by people around 20 years old each year.






\section{Conclusions}
\label{sec:conclusion}

With access to more and more information, it becomes essential to shrink many of these documents to a human-friendly level. For many generations, news providers had done this, who selected the most important news for the general public. Nowadays, however, even the reporters might struggle to work with these endless news feeds.

If it were possible to automate these reporters, it would help millions of people. Imagine a  TV news network, which only needs a host, and AI takes care of the rest. With this approach, it might be even possible to create a fully unbiased news coverage, which many people demand.

Here we proposed automated social media accounts that provide a reliable news feed to their followers. To achieve this, the project uses a modified DBSCAN for clustering. Then the program evaluates each cluster, and if rated important enough, an informative post is published. We believe this project is a step in the right direction of creating a fully unbiased news source and automating social media accounts concerned with news coverage. 


\section{Conflict of Interest}
We wish to confirm that there are no known conflicts of interest associated with this publication and there has been no significant financial support for this work that could have influenced its outcome.






\Urlmuskip=0mu plus 1mu
\bibliographystyle{elsarticle-num} 
\bibliography{citations}




\section*{Current executable software version}


\begin{table}[!ht]
\begin{tabular}{|l|p{6.5cm}|p{6.5cm}|}
\hline
\textbf{Nr.} & \textbf{(Executable) software metadata description} & \textbf{Please fill in this column} \\
\hline
S1 & Current software version & v1.0.0 \\
\hline
S2 & Permanent link to executables of this version  & \url{https://github.com/MichalBravansky/MejPrehled} \\
\hline
S3 & Legal Software License & MIT \\
\hline
S4 & Computing platforms/Operating Systems & Linux, Microsoft Windows \\
\hline
S5 & Installation requirements \& dependencies & \\
\hline
S7 & Support email for questions & michal.bravansky1@gmail.com \\
\hline
\end{tabular}
\caption{Software metadata (optional)}
\end{table}

\end{document}